\documentstyle[12pt,epsf]{article}

\global\arraycolsep=1pt
\def\be{\begin{equation}}
\def\ee{\end{equation}}
\def\bea{\begin{eqnarray}}
\def\eea{\end{eqnarray}}
\def\w{\omega}

\def\gm{\gamma}
\def\lm{\lambda}
\def\ra{\rightarrow}
\def\hf{{1 \over 2}}
\def\ff{{1 \over 4}}
\begin{document}
\begin{flushleft}
{\baselineskip=14pt 
\rightline{
 \vbox{\hbox{YITP-98-1}
       \hbox{hep-th/9801079}
       \vskip 2mm
       \hbox{January 1998}       }}}
\end{flushleft}

\vskip 1.0 cm

{\Large
\centerline{{\bf General Solution of 7D Octonionic Top Equation}}
\vskip 1.5cm

\centerline{Tatsuya Ueno}}
\vskip 1.5cm

\centerline{{\it Yukawa Institute for Theoretical Physics},}
\vskip 5pt
\centerline{{\it Kyoto University},}
\vskip 5pt
\centerline{{\it Kyoto 606-01, Japan.}}
\vskip 5pt
\centerline{\quad{\sl tatsuya@yukawa.kyoto-u.ac.jp}} 

\vskip 3.0cm

\begin{center}
{\bf ABSTRACT}
\end{center}
The general solution of a 7D analogue of the 3D Euler top equation
is shown to be given by an integration over a Riemann surface with 
genus 9. The 7D model is derived from the 8D $Spin(7)$ invariant 
self-dual Yang-Mills equation depending only upon one variable and 
is regarded as a model describing self-dual membrane instantons.
Several integrable reductions of the 7D top to lower target space
dimensions are discussed and one of them gives 6, 5, 4D descendants
and the 3D Euler top associated with Riemann surfaces with 
genus 6, 5, 2 and 1, respectively.

\thispagestyle{empty}
\newpage
\section{Introduction}%
The relevance and the importance of the algebra $\cal O$ of octonions
in physics have been discussed by many authors, together with other
three division algebras $R$, $C$, $\cal H$ of real, complex and 
quaternionic numbers. 
See, for example, \cite{CDFN}\cite{DGT}\cite{KTE} and references 
therein.
\par

In Ref.\cite{CDFN}, the 4D (anti-)self-dual Yang-Mills equation was 
generalized to higher-dimensional linear relations for the field strength 
$F_{\mu\nu}$, which lead to the full Yang-Mills equation, via the 
Bianchi identity. 
Among various examples noted in \cite{CDFN}, an interesting case 
arises in eight dimensions, which is invariant under the transformation 
by a maximal subgroup $Spin(7)$ of the rotation group $SO(8)$ and 
can be expressed by using the octonionic structure constants $c_{ijk}$ 
$(i,j,k = 1,\cdots,7)$,
\be
F_{8i} = \hf \, c_{ijk}\, F_{jk} \ .              \label{8Dsd}
\ee
Recalling the fact that the 4D self-dual equation can be cast in the 
form of (\ref{8Dsd}) with the quaternionic structure constants 
$\varepsilon_{ijk}$ instead of $c_{ijk}$, we recognize
the 8D equation (\ref{8Dsd}) to be a natural generalization of 
the usual 4D self-dual one.
Several solutions of (\ref{8Dsd}) and its 7D relatives
were found in \cite{FNFN}\cite{Iva}\cite{GN} and were then 
used to construct string and membrane solitons 
\cite{HS}\cite{Iva}\cite{GN}.
In recent papers \cite{BKS}\cite{AOS}, the 8D equation has been 
applied to construct a topological Yang-Mills theory on Joyce 
manifolds as an 8D counterpart of the 4D Donaldson-Witten theory
\cite{DW}.
\par

It is also recently discussed that (self-dual) Yang-Mills gauge fields 
depending only upon time play a role in the context of M-theory 
\cite{CFZ}\cite{Fair}\cite{FLPT}\cite{HK}.
In the reduction to 1D world-sheet, (\ref{8Dsd}) is modified to the 
form of the 7D Nahm equations, with the gauge condition $A_8 = 0$,
\be
{d \over dt}A_i (t) = \hf c_{ijk}[A_j(t),\ A_k(t)] \ .
\label{Nahm7}
\ee 
The commutator in the right-hand-side (RHS) can be replaced by the 
Poisson bracket (P.B.) if we take the gauge group of the Yang-Mills 
to be the infinite-dimensional group of area-preserving 
diffeomorphisms $SDiff({\cal M})$ on a 2D surface $\cal M$,  
\be
{d \over dt} A_i (t, \sigma, \tau)
= \hf c_{ijk}\{A_j(t, \sigma, \tau),\ A_k(t, \sigma, \tau )\} \ ,
\label{PB7}
\ee 
where matrix indices in (\ref{Nahm7}) are Fourier-transformed to the 
coordinates $(\sigma, \tau)$ on $\cal M$. 
It may be also worthwhile to consider the Nahm equations defined with the 
Moyal bracket \cite{CFZ}\cite{Fair}, which is essentially equivalent
to the commutator of $SU(N)$ matrices and reduces to the P.B. in the limit 
$N \ra \infty$.
\par

The equation (\ref{PB7}) was proposed as an ansatz for self-dual 
membrane instantons embedded in seven space dimensions \cite{CFZ}.
Iterating (\ref{PB7}), with the identity for $c_{ijk}$,
\be
c_{ikp}c_{jlp} = \delta_{ij}\delta_{kl} - \delta_{il}\delta_{kj}
+ T_{ikjl} \ ,
\ee 
where $T_{ikjl} = {1 \over 3!}\varepsilon_{ikjlhmn}c_{hmn}$, we obtain 
a second order equation for $A_i$,
\be
{d^2\over dt^2} A_i  = - \{\{A_i, A_j\}, A_j\} \ ,
\qquad (j \ {\rm summed})
\label{2nd}
\ee 
which arises from the Lagrangian,
\be
{\cal L} = \hf (\dot{A_i})^2 + \ff \{A_i, A_j\}^2 \ .
\label{lag}
\ee 
The Lagrangian is given by a sum of the square of the equation 
(\ref{PB7}) up to a total derivative term and solutions of (\ref{PB7}) 
satisfy the Bogomol'nyi bound in the theory.
A similar Bogomol'nyi property holds for the commutator case 
(\ref{Nahm7}). 
\par

In Ref.\cite{FU}, Fairlie and the author investigated the integrability 
of the 7D Nahm equations (\ref{Nahm7}) with an ansatz 
$A_i(t) = \w_i(t) \, e_i$ ($i$ not summed), where $e_i$ are 7 unit 
imaginary octonions.
With the use of the ansatz, (\ref{Nahm7}) is reduced to a 7D analogue
of the familiar Euler equation for a 3D top, 
\be
\dot{\w}_i(t) = \hf \, c^2_{ijk} \, \w_j(t) \w_k(t) \ . \label{oceuler}
\ee 
The equation is also obtained from the P.B. case (\ref{PB7}) by 
choosing an appropriate basis of functions on the surface $\cal M$ 
\cite{FLPT} and hence, gives self-dual membrane instantons.
It was shown that the 7D top has 6 independent conservation 
laws to ensure the full integrability of the system \cite{FU},
although it remained to find its explicit integral solutions.
\par

In this paper, we give another simplified proof of the integrability of 
the 7D top and show that its general solution is given by an integration 
over a Riemann surface with genus 9.
We discuss the procedure to reduce the 7D top to lower target space 
dimensions while keeping the integrability and obtain its
lower-dimensional integrable descendants.
A reduction is demonstrated to yield 6, 5, 4D models and the 3D Euler
top corresponding to Riemann surfaces with genus 6, 5, 2 and 1, 
respectively.
\par

\section{3D Euler top equation}%
Before going to the 7D top, let us briefly review the 3D Nahm equations
for self-dual fields in four dimensions, 
\be
{dA_i \over dt}= \hf \varepsilon_{ijk}[A_j,\ A_k] \ .
\label{Nahm}
\ee
The set of non-denegerate linear transformations making (\ref{Nahm})
invariant is the group $SU(2) = Aut({\cal H})$, the group of 
automorphisms on $\cal H$. 
Assuming an ansatz for a solution of (\ref{Nahm}), with a matrix 
representation of quaternions ($SU(2)$), e.g. the Pauli matrices 
$\sigma_i$,
\be
A_i = \w_i\, \sigma_i\ , \qquad  (i\ {\rm not~~summed})
\label{ans3}
\ee
we obtain the standard 3D (Euclidean) Euler top equations, 
\be
\dot \w_1 =\w_2 \, \w_3 \ , \qquad
\dot \w_2 =\w_3 \, \w_1 \ , \qquad
\dot \w_3 =\w_1 \, \w_2 \ .
\label{euler3}
\ee
The system is integrable, for we have two independent conserved 
quantities,
\be
\w_1^2 - \w_2^2 = c_2 \ , \ \ \ \ \w_1^2 - \w_3^2 = c_3 \ .
\label{cq3}
\ee
Solving (\ref{cq3}) for $\w_2$ and $\w_3$ and substituting them into 
the equation for $\w_1$, we have
\be
\dot{\w}_1 = \sqrt{(\w_1^2 - c_2)(\w_1^2 - c_3)} \ .
\label{elliptic}
\ee
As is well known, the Riemann surface associated with the integration 
of (\ref{elliptic}) is of genus $g=1$ (torus) and the general solution
of the 3D top is given by elliptic functions.
\par

We note that the set of symmetry transformations of (\ref{euler3})
becomes a finite group with 24 elements, unlike the case of (\ref{Nahm}). 
All the elements of the group are generated by the products of 3
generators changing the sign of two of $\w_i$'s and 6 permutations 
of $\w_i$'s.
Since, in the latter, there appear 3 permutations with determinant 
$-1$, the finite group is not a subgroup of $SU(2)$. 
\par

\section{7D top equation}%
The multiplication rule in the algebra $\cal O$ is specified by the 
relation among the octonions,
\be
 e_i \, e_j = - \delta_{ij} 1 + c_{ijk} \, e_k \ .  \label{octo}
\ee
The exceptional group $G_2$ is nothing but the group $Aut({\cal O})$ 
of automorphisms on $\cal O$, which preserve the relation (\ref{octo}).
The 7D Nahm equations (\ref{Nahm7}) for 8D self-dual fields are
invariant under the linear transformation by the group $G_2$.
\par

In this paper, we take the following explicit realization of the totally 
anti-symmetric $c_{ijk}$,
\be
c_{127}=c_{631}=c_{541}=c_{532}=c_{246}=c_{734}=c_{567}=1 \ , 
\qquad
(\rm{ others \ zero})
\ee
then the relation for $e_i$ can be read off diagrammatically from Fig.1; 
$e_1e_2 = e_7 = - e_2 e_1$, $e_2e_4 = e_6 = - e_4e_2$ etc..
The diagram, the seven-point plane, arises from the projective geometry of 
a plane over a finite field of characteristic two; 3 points lie  on
every line and 3 lines pass through each point. 
The 7D top equation (\ref{oceuler}) becomes, with this choice of $c_{ijk}$,
\bea
&&
{d \over dt}\w_1= \w_2\w_7+\w_6\w_3+\w_5\w_4 \ ,
\quad
{d \over dt}\w_2=\w_7\w_1+\w_5\w_3+\w_4\w_6 \ ,
\nonumber\\
&&{d \over dt}\w_3=\w_1\w_6+\w_2\w_5+\w_4\w_7 \ ,
\quad
{d \over dt}\w_4=\w_1\w_5+\w_6\w_2+\w_7\w_3 \ ,
\label{euler7}\\
&&{d \over dt}\w_5=\w_4\w_1+\w_3\w_2+\w_6\w_7 \ ,
\quad
{d \over dt}\w_6 =\w_3\w_1+\w_2\w_4+\w_7\w_5 \ ,
\nonumber\\
&&{d \over dt}\w_7=\w_1\w_2+\w_3\w_4+\w_5\w_6 \ .
\nonumber
\eea
The contributions to $\dot \w_i$ come from the products of the pairs 
of $\w$'s associated with the other points on each of the three lines 
through $i$ in Fig.1.
\par

The $G_2$ invariance of the 7D Nahm equations (\ref{Nahm7}) breaks
down after the use of the ansatz $A_i(t) = \w_i(t) \, e_i$ and as 
in the 3D case, the symmetry group of the 7D top (\ref{euler7})
becomes finite with 7 generators changing the sign of all variables 
except for the three $\w_i$'s which are associated with the points on 
each of seven lines and 168 permutations keeping the relative
structure of points in Fig.1, that is, the permutations by which three 
points on a line are transformed to those on a line.
\par

\begin{figure}[b]
\epsfxsize= 45 mm
\begin{center}
\leavevmode
\epsfbox{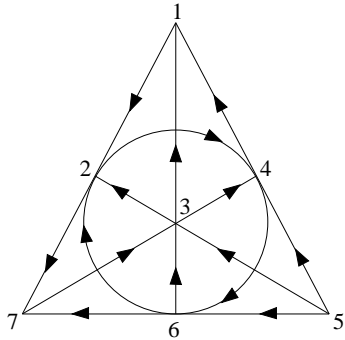}
\end{center}
\caption{7 point plane.}
\end{figure}
It is known that octonions $e_i$ do not have any matrix representation,
due to the lack of associativity.
However their adjoint-like representation given by $7 \times 7$ matrices
$(C^i)_{jk} = c_{ijk}$ with
\be
C^i = {1 \over 6} c_{ijk}[C^j, C^k] \ ,
\qquad \ \ 
{\rm Tr} C^i C^j = - 6 \delta_{ij} \ ,
\label{cc}
\ee
works to give the 7D top equation, up to a factor $\hf$ in the RHS of 
(\ref{oceuler}), via the ansatz $A_i = \w_i \, C^i$.
A P.B. equation similar to the first one in (\ref{cc}) was given for a 
basis of functions $f_i(\sigma, \tau)$ on a surface $\cal M$ in the 
context of membrane instantons \cite{FLPT}.
\par

A type of special solutions of the 7D top can be obtained by
considering the embedding of the 3D Euler top into the 7D one.
{}From the structure of (\ref{euler7}), we can recognize that there
are 7 ways of embedding, each of which takes three variables
$(\w_i,\w_j,\w_k)$ on a line in Fig.1 and makes the other four 
variables zero. 
Then the seven equations (\ref{euler7}) reduce to the 3D top 
equations (\ref{euler3}).
\par

\subsection{Integrability}
In order to prove the integrability of the 7D top, it is convenient to
define the following linear combinations $a_i(t)$ of seven variables 
$\w_i(t)$, 
\bea
&&a_1 = \w_3+\w_4+\w_5+\w_6 \ , \qquad a_2 = \w_1+\w_2+\w_5+\w_6 \ ,
\nonumber \\
&&a_3 = \w_1+\w_3+\w_5+\w_7 \ , \qquad a_4 = \w_2+\w_4+\w_5+\w_7 \ ,
\label{abc} \\
&&a_5 = \w_2+\w_3+\w_6+\w_7 \ , \qquad a_6 = \w_1+\w_4+\w_6+\w_7 \ ,
\nonumber \\
&&a_7 = \w_1+\w_2+\w_3+\w_4 \ , 
\nonumber 
\eea
then the equations of motion (\ref{euler7}) become
\be
\dot{a}_i = \ff \, a_i \, (\sum_{j=1}^7 a_j - 4 a_i) \ ,
\qquad \ \ (i=1,\cdots,7)
\label{eqabc}
\ee
which give the time derivatives for the difference of $a_i$'s,
\be
\dot{(a_i - a_k)} = \ff \, (a_i - a_k) \, 
(\sum_{j=1}^7 a_j - 4 a_i - 4 a_k) \ .
\label{eqaa}
\ee
We introduce a quantity $W$ with constants $\rho_i$ and
$\chi_{ij}$,
\be
W = \sum_i \rho_i \ln{a_i} + \sum_{i < j}\chi_{ij} \ln{(a_i - a_j)}
\ ,
\ee
then its time derivative is 
\be
\dot{W} = \ff \{(\sum_i \rho_i + \sum_{i<j}\chi_{ij}) \sum_k a_k 
- 4 (\sum_i \rho_i a_i + \sum_{i<j} \chi_{ij}(a_i + a_j)) \} \ .
\label{tW}
\ee
The condition $\dot{W} = 0$ requires that the coefficients of $a_i$'s in 
the RHS of (\ref{tW}) should be zero.
These seven constraints for $\rho_i$ and $\chi_{ij}$ can be solved for
$\rho_i$ and $W$ is expressed in terms of 21 constants of motion $N_{ij}$,
\be
W = \sum_{i<j} \chi_{ij} \ln{N_{ij}} \ , \qquad \ \ \ \ \ 
N_{ij} = (\prod_{k=1}^7 a_k)^{1 \over 3} (a_i - a_j)/a_ia_j \ ,
\label{nij}
\ee
although not all of $N_{ij}$ are independent,
\be
N_{ij} = N_{1j} - N_{1i} \ .
\label{nnn}
\ee
This relation shows that the set of six $N_{1j}$ $(j \not=1)$ becomes
a basis of conserved quantities in the 7D top.
\par

We can prove $N_{1j}$ to be functionally independent, which ensures
the full integrability of the system.
In \cite{FU}, we defined 7 conserved quantities $\gm_i$, being
quartic polynomials in $\w_i$'s,
\be
\gm_i = N_{j_1k_1}N_{j_2k_2}N_{j_3k_3} 
= a_i(a_{j_1}-a_{k_1})(a_{j_2}-a_{k_2})(a_{j_3}-a_{k_3}) \ , 
\ee
where $(j_p,k_p)$, $(j_p < k_p, \ p = 1,2,3)$ lie on the respective three 
lines through the point $i$.
We found that there exists only one constraint for the 7 $\gm_i$'s, 
showing the independence of $N_{1j}$.
Here, let us check the independence by solving the second equations in 
(\ref{nij}) for $a_j$ in terms of a variable, say, $a_1$.
Defining the constants of motion $(\lm_i, \xi_i)$, given by the initial 
values $a_{i0}$ of $a_i$,
\be
\lm_i = N_{1i}/N_{12} = a_{20} (a_{10} - a_{i0})/a_{i0} (a_{10} - a_{20}) 
 \ , \qquad \ \ \xi_i = 1 - \lm_i \ ,
\ee
we can express $a_i$ in terms of $a_1$ and $a_2$,
\be
a_i = a_1 a_2/(\lm_i a_1 + \xi_i a_2) \ ,
\label{ai}
\ee
where $\lm_1 = 0$ and $\lm_2 = 1$.
The variable $a_2$ is expressed implicitly as a function of $a_1$
through the relation obtained by substituting (\ref{ai}) into the 
equation for $N_{12}$ in (\ref{nij}),
\be
{(a_1a_2)^3(a_1-a_2)^3 \over (\lm_3 a_1 + \xi_3 a_2)
(\lm_4 a_1 + \xi_4 a_2)(\lm_5 a_1 + \xi_5 a_2)(\lm_6 a_1 + \xi_6 a_2)
(\lm_7 a_1 + \xi_7 a_2)} = N \ ,
\label{bc2}
\ee
where $N$ is a constant,
\be
N = (N_{12})^3 = a_{30}a_{40}a_{50}a_{60}a_{70} (a_{10}-a_{20})^3 / 
a_{10}^2 a_{20}^2 \ .
\label{N}
\ee
\par

\begin{figure}[t]
\epsfxsize= 100 mm
\begin{center}
\leavevmode
\epsfbox{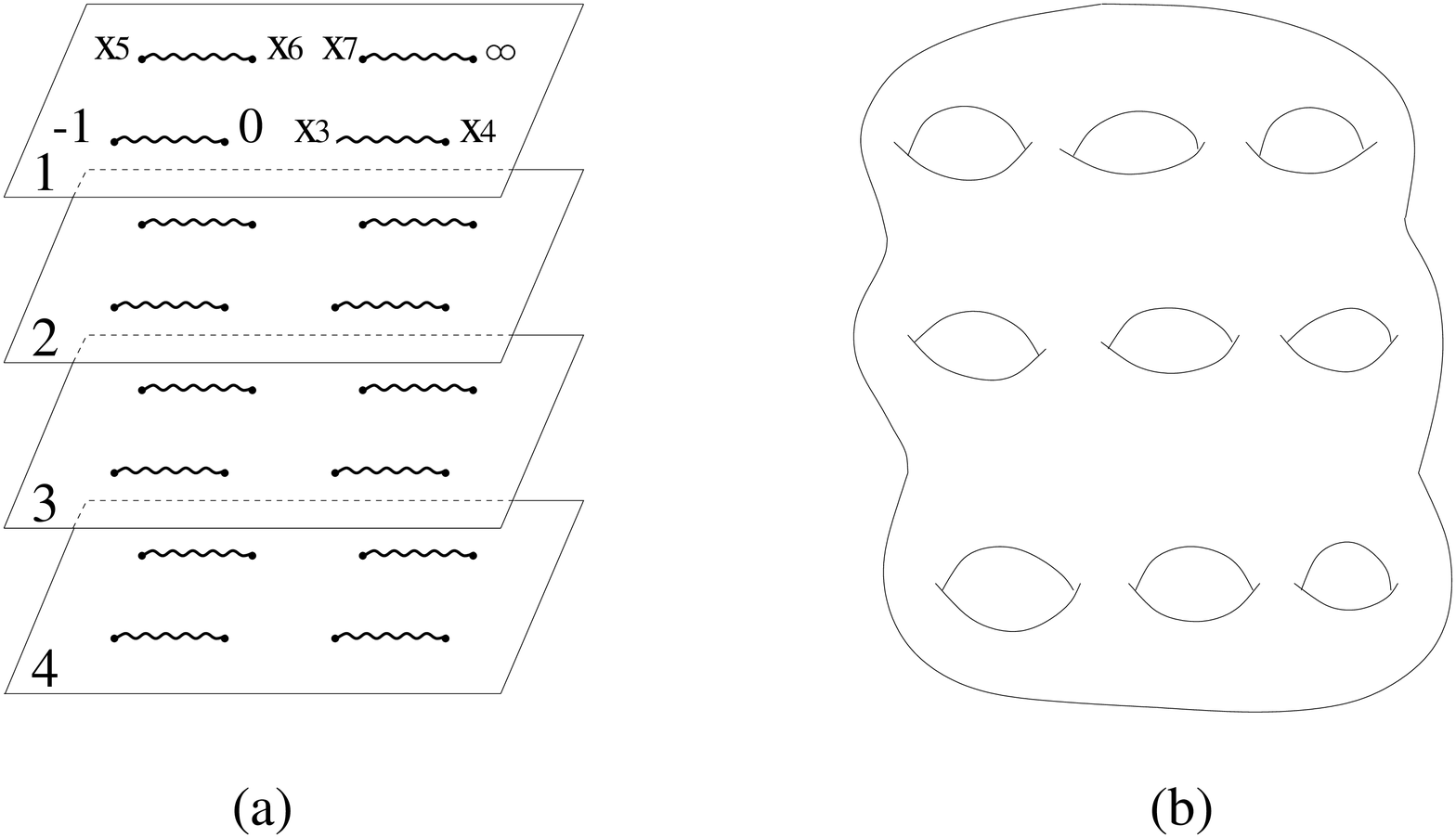}
\end{center}
\caption{Riemann surface associated with the curve
$y^4 = N R(R+1)\prod_{p=3}^7(\xi_pR-\lm_p)$ for the general solution 
of the 7D top. In (a), $x_p = \lm_p/\xi_p$.}
\end{figure}

\subsection{General Solution}
We introduce the ratio of $a_2$ and $a_1$, $R(t) = - a_2(t)/a_1(t)$ and 
obtain the equation of motion for $R(t)$ using (\ref{eqaa}),
\be
\dot{R} = a_1 R(R+1) \ . 
\label{Rceq}
\ee
Solving (\ref{bc2}) for $a_1$ in terms of $R$, we have
\be
a_1^4 = (a_1(R))^4 = N {(\xi_3 R - \lm_3)(\xi_4 R - \lm_4)
(\xi_5 R - \lm_5)(\xi_6 R - \lm_6)(\xi_7 R - \lm_7)
\over R^3(R+1)^3} \ ,
\label{cR}
\ee
and the other variables are expressed in terms of $R$ as
\be
a_i = a_1(R)R/(\xi_i R - \lm_i) \ .
\label{Rabc} 
\ee
Using (\ref{Rceq}) and (\ref{cR}), we obtain a first-order 
equation for the ratio $R$,
\be
{\dot R} = {\root 4 \of {N R (R+1)(\xi_3 R - \lm_3)(\xi_4 R - \lm_4)
(\xi_5 R - \lm_5)(\xi_6 R - \lm_6)(\xi_7 R - \lm_7)}} \ .
\label{Req}
\ee
The integral associated with this equation can be shown to 
correspond to a Riemann surface with $g=9$ in Fig.2(b); 
the order $1/4$ in the RHS of (\ref{Req}) means that we 
need four complex surfaces, each of which has four cuts 
since the order of $R$ is 7 in the RHS.
Picking up a cut on each surface in Fig.2(a), attaching the total of
four cuts together and repeating it for other three sets of four cuts, 
we have a Riemann surface with 9 handles.
\par

\begin{figure}[ht]
\epsfxsize= 150 mm
\begin{center}
\leavevmode
\epsfbox{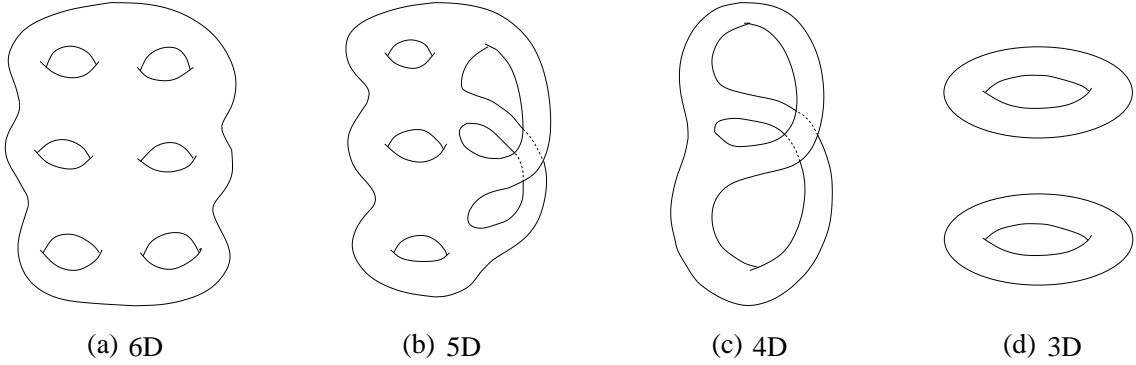}
\end{center}
\caption{Riemann surfaces for lower-dimensional descendants
of the 7D top.}
\end{figure}
\subsection{Integrable reduction to lower dimensions}
Let us consider the reduction of the 7D top to lower target space
dimensions which is compatible with the equations of motion 
and hence preserves the integrability.
In the case that the variables $a_i$ and $a_j$ are not independent, 
that is, $a_i = r a_j$ with a constant $r$, only two cases 
$r=0, 1$ are permitted from (\ref{eqaa}),
\be
a_i = 0 \ ,  \qquad \qquad a_i = a_j \ .   \label{rd}
\ee
Imposing the conditions to (\ref{eqabc}), we obtain lower-dimensional 
integrable descendants of the 7D top successively.
In the section 3.2, it is assumed implicitly that $a_1$ and $a_2$ 
are independent and hence we impose (\ref{rd}) to the other five 
variables $a_p$ $(p=3,4,5,6,7)$.
Then we encounter the following four cases; \\
(i) $a_p \ra a_q$;
from the definition of the constants $(\lm_p,\xi_p)$, two factors 
in (\ref{Req}) coincide with each other,
\be
(\xi_p R - \lm_p)(\xi_q R - \lm_q) \ \ra \
(\xi_p R - \lm_p)^2 \ .
\ee
(ii) $a_p \ra 0$;
in this limit, $(\lm_p,\xi_p)$ become singular but the 
products of them and the constant $N$ tend to 
\be
N\, (\xi_p R - \lm_p) \ \ra \
- M_p (R + 1) \ ,  \qquad 
M_p = \prod_{q=3,\not=p}^7 a_{q0} 
\, (a_{10}-a_{20})^2/a_{10}a_{20} \ .
\ee
(iii) $a_p \ra a_1$; 
the pair $(\lm_p,\xi_p)$ tends to $(0,1)$ and
$(\xi_p R - \lm_p) \ \ra \ R$.
\\ 
(iv) $a_p \ra a_2$;
the pair $(\lm_p,\xi_p)$ tends to $(1,0)$ and
$(\xi_p R - \lm_p) \ \ra \ - 1$.
\\
The procedures (i),(ii),(iii) give a square factor in the RHS of 
(\ref{Req}), while a factor in the RHS disappears under (iv).
\par

We shall show a reduction of the 7D top to lower-dimensional
descendants, whose corresponding Riemann surfaces are indicated in 
Fig.3. 
\\
(a) 6D case; in the limit (iv) $a_3 \ra a_2$, the factor 
$(\xi_3 R - \lm_3)$ in (\ref{Req}) goes to $-1$.
Then the four cuts on each surface in Fig.2(a) reduce to three cuts,
and by the same procedure as in the 7D case, we find that the Riemann
surface for the 6D model is of genus $g=6$.
\\
(b) 5D case; we take the limit (i) $a_6 \ra a_7$, then 
(\ref{Req}) becomes
\be
{\dot R} = {\root 4 \of {-N R (R+1)(\xi_4 R - \lm_4)
(\xi_5 R - \lm_5)(\xi_7 R - \lm_7)^2}} \ .
\label{Req5}
\ee
In Fig.2(a), we set three cuts between points $(-1,0)$, 
$(\lm_4/\xi_4, \lm_5/\xi_5)$ and $(\lm_7/\xi_7, \infty)$.
The third cut $(\lm_7/\xi_7, \infty)$ has a different property from 
the others; an orbit passing the cut in the surface 1 goes to the 
surface 3, while the orbit in the surface 2 goes to 4.
Copies of the cut in the surfaces 1 and 3 and those in 2 and 4 have to
be identified separately, which produces two handles 
drawn as overlapped ones in Fig.3(b) and hence $g=5$.
\\
(c) 4D case; in the limit (ii) $a_5 \ra 0$, (\ref{Req5}) reduces to
\be
{\dot R} = {\root 4 \of {M_5 R (R+1)^2 (\xi_4 R - \lm_4)
(\xi_7 R - \lm_7)^2}} \ .
\label{Req4}
\ee
We set two cuts between points $(0, \lm_4/\xi_4)$ and
$(-1, \lm_7/\xi_7)$ where copies of the latter in the surfaces 1 and 3 
and those in 2 and 4 are identified separately, which gives a $g=2$
surface.
\\
(d) 3D Euler top; (\ref{Req4}) becomes under the limit (iii) 
$a_4 \ra a_1$,
\be
{\dot R} = {\root 4 \of {M_5 R^2 (R+1)^2 (\xi_7 R - \lm_7)^2}} 
= \pm {M_5}^{1 \over 4} \sqrt{R(R+1)(\xi_7 R - \lm_7)} \ .
\label{Req3}
\ee
The power of the RHS reduces to $1/2$ and the equation (\ref{Req3}) 
describes the 3D Euler top in the section 2 defined with three 
variables $(\w_1,\w_4,\w_5)$. 
The pairs of surfaces (1,3) and (2,4) are disconnected from each other 
and represented by two tori in Fig.3(d), each of which 
corresponds to the sign $+$ or $-$ in the RHS of (\ref{Req3}).
\\
(e) 2D case; taking $a_7 \ra 0$ or $a_1$, (\ref{Req3}) goes to a
descendant equation associated with a torus in which one of two 
cycles is shrunk. Another limit $a_7 \ra a_2$ gives a descendant
corresponding to a sphere. 
\par

Adding to the above reduction, there are various other ways of reducing 
the 7D top by the successive use of the procedures (i) to (iv).
Some of them may yield higher order factors in (\ref{Req}).
For example, a reduction of the 7D top to a 5D model with 
$a_5 = a_6 = a_7$ gives a cubic factor,
\be
{\dot R} = {\root 4 \of {N R (R+1)(\xi_3 R - \lm_3)(\xi_4 R - \lm_4)
(\xi_7 R - \lm_7)^3}} \ .
\label{R3}
\ee
In this case, we can set three cuts $(-1,0)$, $(\lm_3/\xi_3, \lm_4/\xi_4)$
and $(\lm_7/\xi_7, \infty)$ in each of the surfaces in Fig.2(a).
The last cut arising from the cubic factor is different from the others in 
the sense that an orbit passing surfaces $1 \ra 4 \ra 3 \ra 2$ via one
of the former two cuts goes to the opposite direction 
$1 \ra 2 \ra 3 \ra 4$ through the last cut.
The last cut, however, plays the same role as the other two when we 
construct the surface for the 5D model, which is of genus $g=6$.
It is also possible to make quartic or more than quartic factors in
(\ref{Req}), although they generate singular points in the 
corresponding Riemann surfaces.
A sort of singular points also arise from the case with lower order 
factors; for example, in a reduction of the 7D top to a 6D model by 
$a_7 \ra a_6$, we have to use the point $\infty$ twice to make four
cuts.
The point $\infty$ may become singular and would not produce a proper
Riemann surface.
\par

To complete the discussion in this section, we consider the 1D reduction 
of the 7D top.
The equation (\ref{Req}) is not available for this case since $a_1$ and 
$a_2$ are now not independent.
As noted around (\ref{rd}), in the 1D case, each of the variables 
$a_i$ has to be zero or equal to the other variables. \\
1. All $a_i$ are equal; substituting them into (\ref{abc}),
(\ref{eqabc}) directly, we have 
\be
\w_i = - {1 \over 3t} \ .
\label{1d1}
\ee
2. One of $a_i$'s is 0;
\be
\w_i = \w_j = \w_k = - {1 \over t} \ , \qquad {\rm others} \ \, 0 \ ,
\label{1d2}
\ee
where $(i,j,k)$ are on each line in Fig.1. 
These are solutions of the 3D top equation.
\\
3. Two of $a_i$'s are 0;
\be
\w_i = - {3 \over t} \ , \qquad \w_j = \w_k = {1 \over t} \ , 
\qquad {\rm others} \ \, - {1 \over t} \ ,
\label{1d3}
\ee
where points $(j,k)$ lie on the respective three lines through the
point $i$. \\
4. Three of $a_i$'s are 0; 
\be
\w_n = c \ \, ({\rm const.}) \ , \qquad {\rm others} \ \, 0 \ ,
\label{1d4a}
\ee
and 
\be
\w_i = - {c \over 2} \ , \qquad \ \ \w_j = \w_k = \w_l = 0 \ ,
\qquad {\rm others} \ \, {c \over 2} \ ,
\label{1d4b}
\ee
where $(j,k,l)$ lie on each of the four lines which does not connect 
to the point $i$. 
\par

All other solutions are obtained from the above ones 
(\ref{1d1}),(\ref{1d2}),(\ref{1d3}) acted upon by the seven sign-changing 
transformations noted in the beginning of the section 3.
\par

\newpage
\section*{Acknowledgements}
I am grateful to D.B. Fairlie for various discussions and suggestions. 
I also thank R. Sasaki, T. Oota and R. Kubo for helpful discussions.
This work is supported by the Japan Society for the Promotion of 
Science, No.\,6293.
\vskip 1.8cm 



\begin{thebibliography} {9}
\bibitem{CDFN}{E. Corrigan, C. Devchand, D.B. Fairlie and J. Nuyts,
              Nucl.~Phys.~{\bf B214} (1982) 452.}

\bibitem{DGT}{R. D\"undarer, F. G\"ursey and C-H. Tze,
J.~Math.~Phys.~{\bf25} (1984) 1496. \\
M. G\"unaydin and F. G\"ursey, J.~Math.~Phys.~{\bf 14}
(1973) 1651.}

\bibitem{KTE}{T. Kugo and P. Townsend, Nucl.~Phys.~{\bf B221}
(1983) 357. \\
J.M. Evans, Nucl.~Phys.~{\bf B298} (1988) 92.}

\bibitem{FNFN}{D.B. Fairlie and J. Nuyts, 
J.~Phys.~A:Math.~Gen.~{\bf 17} (1984) 2867. \\
S. Fubini and H. Nicolai, Phys.~Lett.~{\bf B155} (1985) 369.}

\bibitem{Iva}{T.A. Ivanova, Phys.~Lett.~{\bf B315} (1993) 277.}

\bibitem{GN}{M. G\"unaydin and H. Nicolai, Phys.~Lett.~{\bf B351} (1995) 
169.}

\bibitem{HS}{J.A. Harvey and A. Strominger, Phys.~Rev.~Lett.~{\bf 66} 
(1991) 549.}

\bibitem{BKS}{L. Baulieu, H. Kanno and I.M. Singer, `Special Quantum
Field Theories In Eight And Other Dimensions' (1997) 
{\bf hep-th/9704167}, PAR--LPTHE 97/07.}

\bibitem{AOS}{B.S. Acharya, M. O'Loughlin and B. Spence, 
Nucl.~Phys.~{\bf B503} (1997) 657.}

\bibitem{DW}{S.K. Donaldson, Topology~{\bf 29} (1990) 257. \\
E. Witten, Commun.~Math.~Phys.~{\bf 117} (1988) 353.}

\bibitem{CFZ}{T. Curtright,  D.B. Fairlie and C.K. Zachos,
Phys.~Lett.~{\bf B405} (1997) 37.}

\bibitem{Fair}{D.B. Fairlie, `Moyal Brackets in M-Theory' (1997)
{\bf hep-th/9707190}, YITP-97-40.}

\bibitem{FLPT}{E.G. Floratos and G.K. Leontaris,` Octonionic Selfduality
for SuperMembranes' (1997) {\bf hep-th/9710064}. \\
E.G. Floratos, G.K. Leontaris, A.P. Polychronakos and R. Tzani, 
`On the Instanton Solutions of the Self-dual Membrane' (1997)  
{\bf hep-th/9711044}, CERN-TH/97-311, IOA-TH.97-15, UUTP-22/97.} 

\bibitem{HK}{S. Hirano and M. Kato, `Topological Matrix Model' (1997) 
{\bf  hep-th/9708039}, OU-HET274, UT-Komaba/97-11.}

\bibitem{FU}{D.B. Fairlie and T. Ueno `Higher-dimensional Generalisations 
of the Euler Top Equations' (1997) {\bf hep-th/9710079}, YITP-97-48,
DTP-97-53.}
\end{thebibliography}
\end{document}